# Atomistic study of electrostatics and carrier transport properties of CNT@MS$_2$ (M= Mo,W) and CNT@BN core-shell nanotubes

Amretashis Sengupta[1,2*]

[1]*Hanse-Wissenschaftskolleg (HWK), Lehmkuhlenbusch 4, 27753 Delmenhorst, Germany*

[2]*Bremen Center for Computational Materials Science (BCCMS), Universität Bremen, Am Fallturm 1, 28359 Bremen, Germany*

**E-mail:* sengupta@bccms.uni-bremen.de

***Abstract:*** *In this work we present an ab-initio study of electronic properties of 1 dimensional (1D) core-shell nanostructures made of MS$_2$ (MoS$_2$, WS$_2$) or BN armchair nanotube encapsulated carbon nanotubes (CNT). With local density approximation (LDA) in density functional theory (DFT) we calculate the bandstructure, carrier effective masses, various fundamental electrostatic features and optical absorption in such core-shell tubes. The carrier transport in these structures are important for nanoelectronics applications and are studied with the Green's function formalism. Simulations show a moderate indirect band gap in the core-shell CNT@MS$_2$ tubes while the CNT@BN shows metallic nature. The varying chirality of CNT strongly affects the carrier effective masses of the CNT@MS$_2$ structure. Electron density is found to be much more localized near the atom cores and stronger in magnitude for the CNT@BN while the W atoms show a more prominent electron-gas presence around them than Mo atoms as found in the electron localization functions. In the CNT@MS$_2$ systems the electrostatic difference potential indicates a drive to transfer charge from the metal to the S atoms in the shell. In terms of optical absorption a strong and sharper peak is observed around 6 eV for the CNT@BN compared to a more broad absorption spectra of the CNT@MS$_2$. Metallic transmission spectra is seen for CNT@BN while CNT@MS$_2$ shows non-metallic transport but with a larger number of transmission states near fermi level. The electronic and optical properties and its possible tuning in the core-shell structures can be useful in various applications such as shielded interconnects, logic switches and optoelectronics.*

## I. INTRODUCTION

Of late nanotubes of non-graphitic layered materials like MoS$_2$, WS$_2$ and hexagonal BN has seen more importance as subject matter of experimental and computational studies. [1]-[8] Nanotubes of MS$_2$ (M=Mo,W) possess a semiconducting band gap with possibility to tune the nature (direct/ indirect) and magnitude of the gap. [1]-[7] Among the new 1D materials MS$_2$ nanotubes also show good stability and optical absoption characteristics, but suffer from a low carrier mobility. [1]-[7] Nanotubes made of hexagonal BN have the advantage of a wide band-gap, high chemical and thermal stability, good thermal conductance and mechanical strength close to that of CNTs. [8]-[11]

Of late MoS$_2$ /CNT composite nanostructures synthesized experimentally have shown great promise for optoelectronics applications, and the fabrication of such MoS$_2$ encapsulated CNT structures have become quite advanced. [12]-[14] Such structures amalgamate the high carrier mobility of CNTs and the versatile optical properties of the MoS$_2$. Also in Li-ion batteries they provide good advantage in terms of high surface area and more versatile chemistry of the MoS$_2$ leading to better functionalization. [6,15]

However WS$_2$ nanotube / CNT hybrid structures have not been as much investigated as their MoS$_2$ counterparts. WS$_2$ having a better carrier transport properties [16] than MoS$_2$ therefore requires more investigation in core-shell structure with CNT.

Wide band gap BN can be provider of good insulation to CNTs. CNT embedded in bulk hBN structures have shown to provide enhanced conductance and higher thermal conductivity than those embedded in SiO$_2$. [17] In this context BN nanotube encapsulated CNT core-shell structures can be particularly useful for interconnect applications in microelectronics industry. CNT@BN nanotubes can be used for providing better insulation and thermal stability to CNT interconnects and at the same time preserving the 1 D nature of the connections.

In this work we carry out ab-initio simulations on MS$_2$ (M=Mo, W) armchair nanotube encapsulated CNT (hereafter referred to as CNT@MS$_2$) and hBN armchair nanotube encapsulated CNT (CNT@BN) structures for their various electronic properties and carrier transport. We consider metallic CNTs of increasing diameter, namely (4,4), (5,5) and (6,6) CNTs that fit well inside (10,10) MS$_2$ NTs or (15,15) BN NTs, without disturbing the structural integrity of the encapsulating tube. With density functional theory (DFT) simulations in QuantumWise AtomistixToolKit (ATK) 2015.1[18], we compute the band-gaps, carrier effective masses, electron density, electron localization function and the electrostatic difference potential in such core-shell tubes. Finally the transmission spectra of the core-shell tubes are calculated atomistically with the Green's function formalism using a Landauer-Buttiker two probe approach. [19]-[25]

## II. METHODOLOGY

Simulations are carried out with local density approximation (LDA) in density functional theory (DFT) [26,27] with Perdew-Zunger (PZ) exchange correlation functional. [26] Troullier-Martins norm-conserving double zeta polarized pseudopotential [27] sets developed by Fritz Haber Institute (FHI-DZP) [28] are used in the calculations. The sets employed are FHI [z=3] DZP for B, FHI [z=4] DZP for C, FHI [z=5] DZP for N, and FHI [z=6] DZP for Mo, W and S. All the sets are inclusive of relativistic core corrections. [29] For simulation of the core-shell structures a 1x1x9 Monkhorst-Pack k-grid [28] is used with cut-off energy of 100 Ha. The self-consistent iterations are controlled by a Pulay mixer algorithm [31], with $10^{-5}$ Ha tolerance and 500 maximum steps. The van der Waals interactions are included by means of Grimme's DFT-D2 method [32] with scale factor of 0.75, cutoff distance 30 Å and damping factor of 20. Default R$_0$ and C$_6$ parameters are taken for the various materials in the DFT-D2 calculations.[32] Geometry optimization for the core-shell supercells is performed with limited memory Broyden-Fletcher-Goldfarb-Shanno (LBFGS) algorithm [33] to a maximum force of 0.05 $eV/Å$ and maximum stress of 0.05 $eV/Å^3$. The optimized cell coordinates are presented in the supplementary information.

Bandstructures are calculated in the $\Gamma - Z$ direction with 1200 points per segment. The carrier effective masses are calculated using a non-parabolic fitting formula [18]

$$m^* = \frac{\hbar^2 k^2}{2E}(1+aE)^{-1} \qquad (1)$$

The valence band maxima and the conduction band minima are fitted to (1) with the local band structure analyser tool in ATK, in the derivative stencil of order 5 with a k-point separation of $0.001 \, Å^{-1}$. The fitting parameter $a$ and RMS deviation for fitting the effective masses (along with the fitting curves) are provided in the Supplementary information.

The self-consistently calculated charge density and that evaluated from the superposition of the wave functions do not necessarily have identical values, and their difference is known as the electron density difference. Physically a positive (negative) value of electron density difference (EDD) on an atom or group of atoms indicates increment (reduction) of charge on that atom. The potential which is created due to this EDD is known as the electrostatic difference potential (EDP). EDP is obtained by solving the Poisson equation for the net charge density difference.

The single particle optical spectra is evaluated from the Kubo-Greenwood formula [34] for the susceptibility tensor. The optical absorption coefficient $\alpha$ is evaluated from the real and imaginary parts of the complex dielectric function ($\varepsilon_r(\omega)$ and $\varepsilon_{im}(\omega)$) as [35,36]

$$\alpha = \sqrt{2}\omega\left\{\sqrt{\varepsilon_r(\omega)^2 + \varepsilon_{im}(\omega)^2} - \varepsilon_r(\omega)\right\}^{1/2} \qquad (2)$$

For calculating the optical spectra a broadening of 0.1 eV is used and 500 bands each above and below the Fermi level are considered for the convergence. 301 points per segment are used in calculation of the spectra from 0 to 10 eV. The equilibrium (zero bias) transmission spectra was obtained from DFT – Green's function (GF) method considering the periodicity of the unit-cell in the c direction. [22,23] For transmission with DFT-GF an even denser 1x1x18 Monkhorst-Pack k-grid is employed, other parameters remaining the same. The Green's function is written as [19]-[23]

$$G(\epsilon) = \left[(\epsilon + i\delta)S - H - \Sigma^r(\epsilon) - \Sigma^\ell(\epsilon)\right]^{-1} \qquad (3)$$

Where $S$ is the overlap matrix and $H$ is the Kohn-Sham Hamiltonian. $\delta$ is an infinitesimally small positive number, $\Sigma^{r,\ell}$ is the right (left) self-energy matrix, related to the system-reservoir coupling $\Gamma$ as

$$\Gamma = i\left(\Sigma - \Sigma^\dagger\right) \qquad (4)$$

The transmission can be calculated from the Green's function as

$$\Im(\epsilon) = \text{Tr}\left[G(\epsilon)\Gamma^{\ell}(\epsilon)G^{\dagger}(\epsilon)\Gamma^{r}(\epsilon)\right] \qquad (5)$$

The fast Fourier transform 2D (FFT2D) Poisson solver is used for the DFT-GF calculations. [24] The self-energy calculator used is Krylov calculator [25], and the energy zero-parameter is the average Fermi level. The transmission spectra is evaluated for the energy range -3 to +3 eV, with 901 points per segment.

## III. RESULTS & DISCUSSION

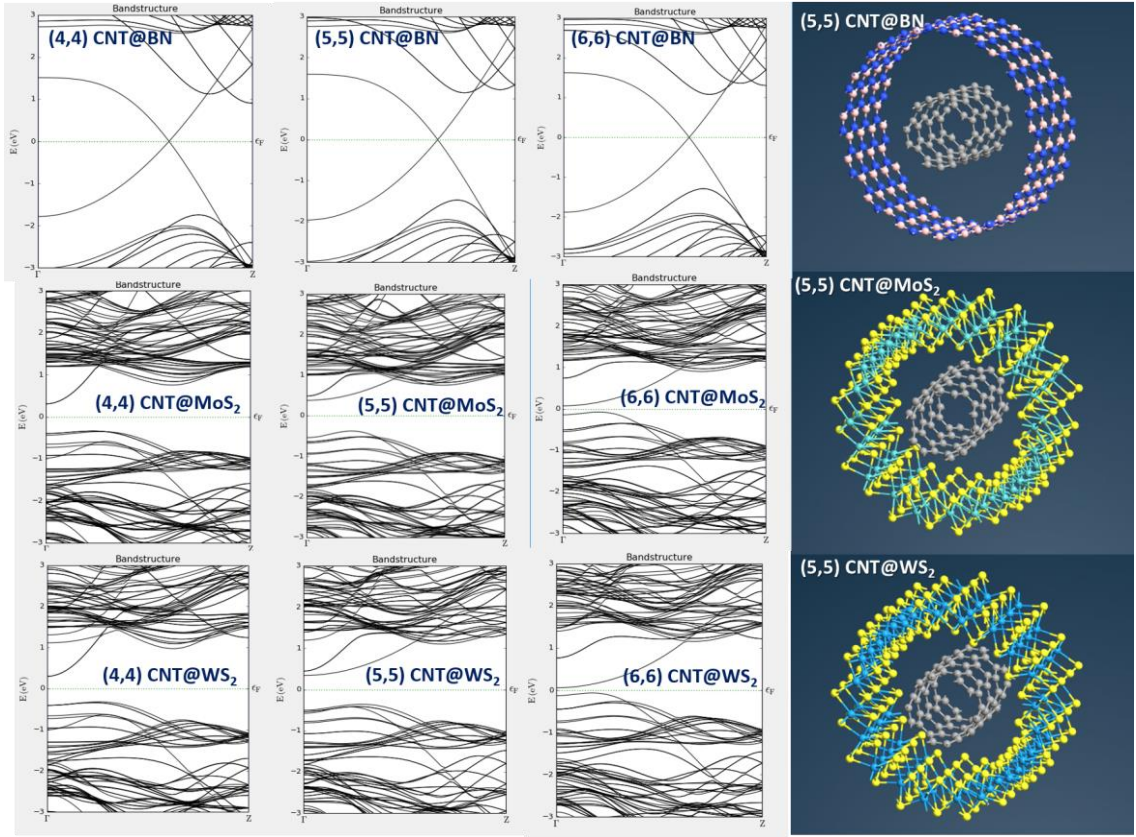

**Fig. 1:** Bandstructures in the Γ-Z direction for the various CNT@MS$_2$ and CNT@BN core-shell nanotubes.core-shell tubes and the schematic supercell

From the DFT calculations, we observe the presence of moderate indirect band gaps for the CNT@MS$_2$ structrures, while metallic nature is seen in the case of CNT@BN. In CNT@MS$_2$ the band gap initially opens up slightly {for (5,5) CNT} and then closes fast {for (6,6) CNT} with the increase in the diameter of the CNT. The gaps retain their indirect nature while closing with conduction band minima (CBM) at the Γ point and the valence band maxima (VBM) at a fractional co-ordinate of (0,0,0.125) (hereafter referred as point P) in the Γ-Z path for the CNT@MS$_2$. For CNT@BN the Dirac point lies a fractional co-ordinate of (0,0,0.315) (hereafter referred as point Q) in the Γ-Z path. With increase in the inner tube diameter, there exist changes both in the number of bands and their interactions/ overlap, with those of the outer tube. The change is more palpable in CNT@MS$_2$ than the CNT@BN owing to slightly smaller inter-tube distances and the strains developed, as presented in Table I and II. The electron and hole effective masses in the axial direction (001) are calculated at the VBM and CBM with a fitting formula in (1). The calculated band gaps and effective masses obtained are listed in Table-I. The details of the fitted effective masses are in supplementary information.

**Table-I:** band gap values and carrier effective masses in the axial direction (001) of the core-shell tubes

| Structure | CNT chirality | Band-gap (eV) | Type of Gap | $m_e(m_0)$ | $m_h(m_0)$ | Inter-tube dist.($\text{Å}$) | c lattice const ($\text{Å}$) |
|---|---|---|---|---|---|---|---|
| **CNT@BN** | (4,4) | metallic | Dirac point at Q | - | - | 7.694 | 2.46100 |
| | (5,5) | metallic | Dirac point at Q | - | - | 6.949 | 2.46100 |
| | (6,6) | metallic | Dirac point at Q | - | - | 6.284 | 2.50399 |
| **CNT@MoS$_2$** | (4,4) | 0.66 | Indirect (Γ-P) | 0.244 | 0.952 | 6.638 | 3.09739 |
| | (5,5) | 0.75 | Indirect (Γ-P) | 1.402 | 0.433 | 5.976 | 3.03514 |
| | (6,6) | 0.16 | Indirect (Γ-P) | 1.476 | 0.579 | 5.149 | 2.99717 |
| **CNT@WS$_2$** | (4,4) | 0.62 | Indirect (Γ-P) | 0.244 | 1.040 | 6.444 | 3.07798 |
| | (5,5) | 0.69 | Indirect (Γ-P) | 1.430 | 0.814 | 5.595 | 3.04039 |
| | (6,6) | 0.12 | Indirect (Γ-P) | 1.445 | 0.879 | 5.053 | 2.99084 |

The effective masses show that in CNT@MS$_2$ we have heavier electrons than holes, except for the (4,4) CNT@MS$_2$ tubes, where holes show heavier effective mass. For CNT@MoS$_2$ there is a big difference in the electron and hole effective masses between (4,4) CNT@MoS$_2$ to (5,5) CNT@MoS$_2$. Although similar change in electron effective mass is observed between (4,4) and (5,5) CNT@WS$_2$, the corresponding change in the hole effective mass is not that staggering. The reason behind such behaviour is the intra-band shifting (among sub-bands) of the VBM and CBM at the Γ and the P points as the chirality of the CNT changes from (4,4) to (5,5). Also variation in straining effects on the nanotubes due to increase in diameter and the stronger overlap of states between the core and the shell can be associated with this change in band-gap and effective mass.

As 2H-MS$_2$ is not lattice matched with 2 D graphene, strain is expected in the core-shell structures. Table –II shows the average tensile/ compressive strain which is calculated in the radial (*rr*) and the axial (*zz*) directions with reference to the CNT/MS$_2$/BN nanotubes in free standing condition and the ones in the core-shell configuration. It is important to mention here that we have to consider the average increase in the diameters as the M and S atoms in MS$_2$ are non-coplanar unlike B and N in hBN and therefore the straining is slightly non-uniform in the radial direction. We take four reference diameters (eight reference points) of the tubes distributed at different angles of 0-360° at 45° intervals. Due to the complex interlayer interactions of the S-M-S units under strain in the outer MS$_2$ shell the inner diameter and outer diameter (essentially the S layers) show more variations at these reference points. The middle layer (i.e. the metal atoms) are thus deemed most fit for the minimizing errors in calculation of the diameters and was thus set as the reference layer for strain calculations. For BN tubes there was negligible difference in the reference diameters but for the $MS_2$ structures variations of the range 1 – 4% were observed in them. The radial strains ($\varepsilon_{rr}$) and the axial strains ($\varepsilon_{zz}$) thus calculated are listed in Table-II.

**Table-II:** Calculated averaged strains on the components of the optimized structures of the core-shell nanotubes

| Structure | CNT chirality | Strain on MS2 / BN NT | | Strain on CNT | |
|---|---|---|---|---|---|
| | | $\varepsilon_{rr}$ (%) | $\varepsilon_{zz}$ (%) | $\varepsilon_{rr}$ (%) | $\varepsilon_{zz}$ (%) |
| **CNT@BN** | (4,4) | +1.729 | -3.370 | -0.644 | +1.301 |
| | (5,5) | +1.677 | -3.269 | -0.335 | +0.674 |
| | (6,6) | +1.699 | -3.313 | -0.519 | +1.041 |
| **CNT@MoS$_2$** | (4,4) | +5.487 | -10.133 | -4.656 | +10.005 |
| | (5,5) | +4.798 | -8.947 | -2.042 | +4.212 |
| | (6,6) | +4.827 | -8.997 | -2.450 | +4.725 |
| **CNT@WS$_2$** | (4,4) | +4.767 | -8.893 | -4.584 | +9.839 |
| | (5,5) | +2.853 | -5.471 | -1.764 | +3.624 |
| | (6,6) | +3.952 | -7.458 | -2.500 | +5.194 |

In all the cases a trend is seen that as the core (CNT) diameter increases, the radial strain on the shell first decreases and then increases slightly again. The change is more prominent for the $MS_2$ shells than the BN shell. In general the strain on the BN NT is much lesser compared to that developed in the $MS_2$ NTs. This is due to the greater amount of separation between the core and the shell in CNT@BN structures and also to some extent the better lattice matching between CNT and BN. The patterns for the radial and axial strain of the CNT show that a radial tensile strain on the shell results in a radial compressive strain whose magnitude has a similar pattern to that of the strain on the shell. However the CNT has much lesser amount of straining than the shell tubes ($MS_2$ or BN). The axial strains are seen to follow a nature reciprocal to the radial strains, which is expected for preserving the structural integrity.

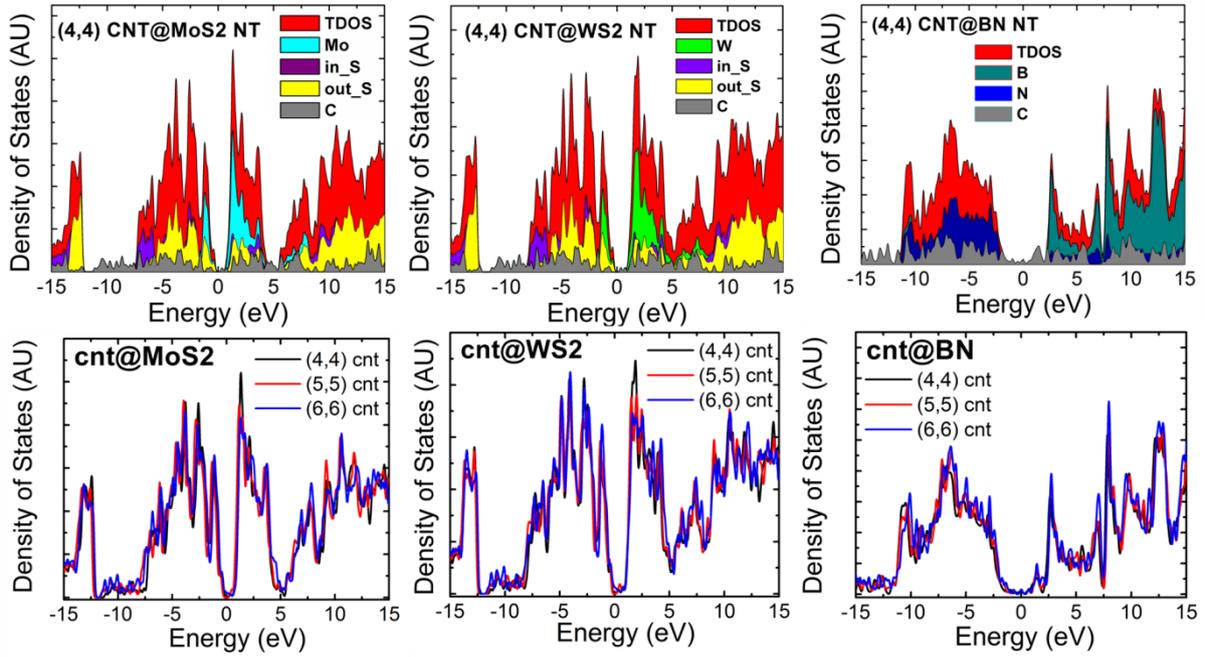

**Fig. 2:** Density of states of the CNT@ MS$_2$ and CNT@BN core-shell nanotubes, top row shows contribution of the elements to the total density of state (TDOS), bottom row only shows the TDOS for the various charalities of the CNT.

With change in the chirality of the CNT the density of states (DOS) (Fig. 2) of the core-shell tubes do not show much significant variation. Slight changes in the peaks near 1.1 eV and 2.25 eV is seen in CNT@MoS$_2$ and CNT@WS$_2$ respectively as chirality varies. For CNT@BN changes of similar magnitude occur further away from the Fermi level (around -5 eV and 7 eV). If we see the contribution of the individual species, we see the contribution towards the total DOS (TDOS) of the system from C atoms is rather less in magnitude than those due to the metal and chalcogen atoms in CNT@MS$_2$ and that due to B and N in CNT@BN tubes. For the CNT@MS$_2$ structure, the contribution of the metal (Mo / W) is the greatest near the Fermi level in the range 5 eV to -1 eV. Sharp peaks of metal states are observed in the said region. The inner and outer layers of S atoms do not provide as significant contribution as metals to the DOS near the Fermi level. The states for the inner and outer S atoms seem to contribute more or less equally in magnitude, but the outer S atom states outweigh the inner S atom states at energies near +10eV and -5 eV. For the range -5 to -7 eV, the inner S atoms provide more states. For the CBT@BN, the B atom states dominate the region above 2.25 eV energy, while in the -2 to +2 eV range only C atom states are present. The contribution of the N atoms is more significant below the Fermi level from energies below -2.25 eV. Overall a higher amount of overlap of bands originating from the inner core and outer shell is observed in case of CNT@MS$_2$, which is quite different from CNT@BN. While a lesser entanglement of states (especially for lower energies) potentially does not offer much variations from the intrinsic properties of the components, in the case of CNT@BN it is a positive factor so as a structure could keep the conducting properties of the CNT interconnect at the same time preventing cross-talk and other isolation issues in nanoelectronic chips.

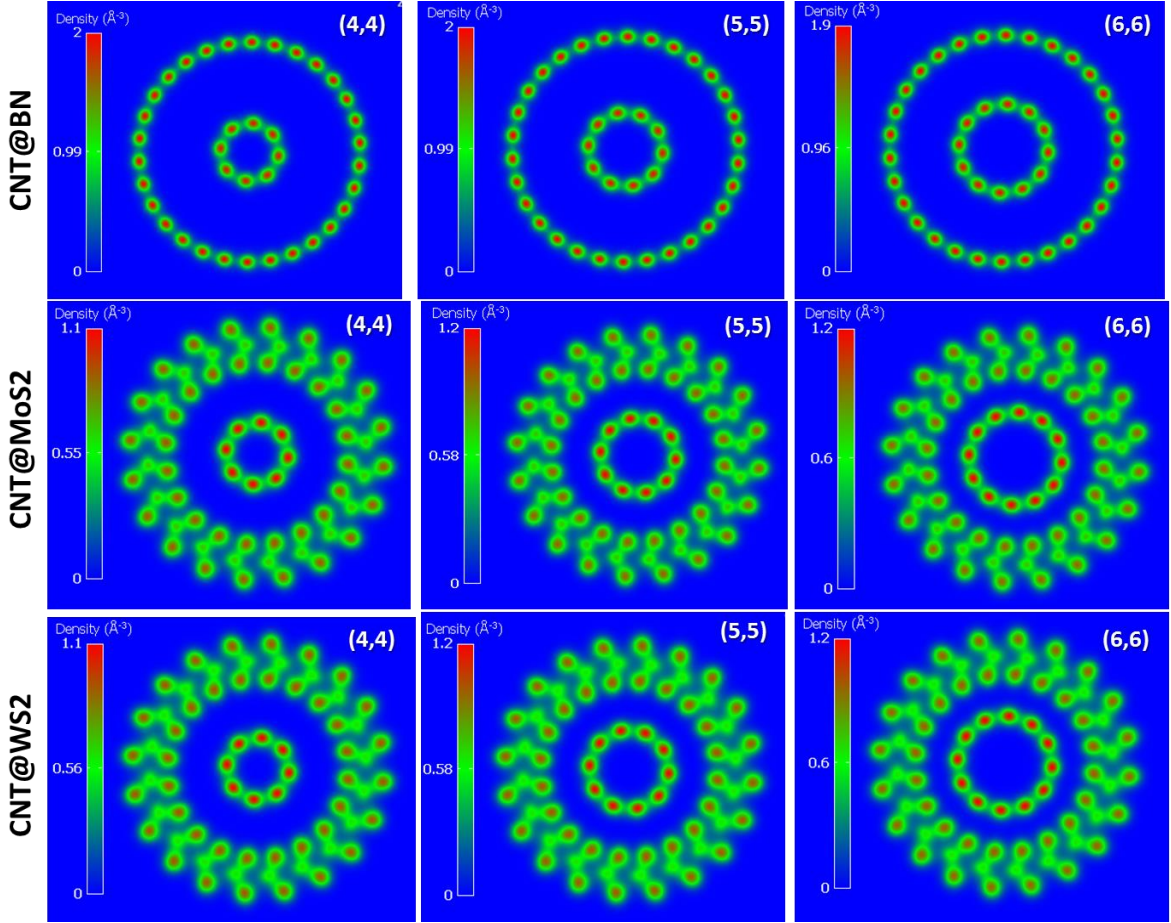

**Fig. 3:** The valence electron density $n(r)$ plots in the x-y cut plane for the CNT@MS$_2$ and CNT@BN core-shell nanotubes. (the type of core-shell NT is indicated on the left side and chirality on the individual tiles.)

Fig. 3 shows the valence electron density $n(r)$ distribution in the x-y cut plane in the core-shell nanotube structures. From the plots we see the electrons to be concentrated mostly around the atom cores in the structures. In the CNT@MS$_2$ tubes the density at the C atoms seem to be maximum, while in the encapsulating MS$_2$ tubes the S atoms show a higher value of $n(r)$ than the metal atoms. Among the two transition metals Mo in comparison to W has a higher electron density around its core, while for the entire MS$_2$ structure, the density appears slightly more smeared in the case of WS$_2$ than MoS$_2$. Magnitude-wise CNT@MoS$_2$ and CNT@WS$_2$ both seem to be of similar electron density. However as we move from (4,4) CNT to (5,5) CNT, in both the cases (of CNT@MS$_2$) we see a slight increase in the charge density (notice the colorbar values in Fig. 2). In comparison the CNT@BN seems to have almost twice the electron density concentrated equally strongly at the atom cores of C, B and N. This trend of concentration of similar charge density between the core and the shell in CNT@BN is unlike CNT@MS$_2$ where the shell atoms had a lesser charge density than the core atoms in general. In case of (6,6) CNT@BN a minute decrease in the charge density is noticeable.

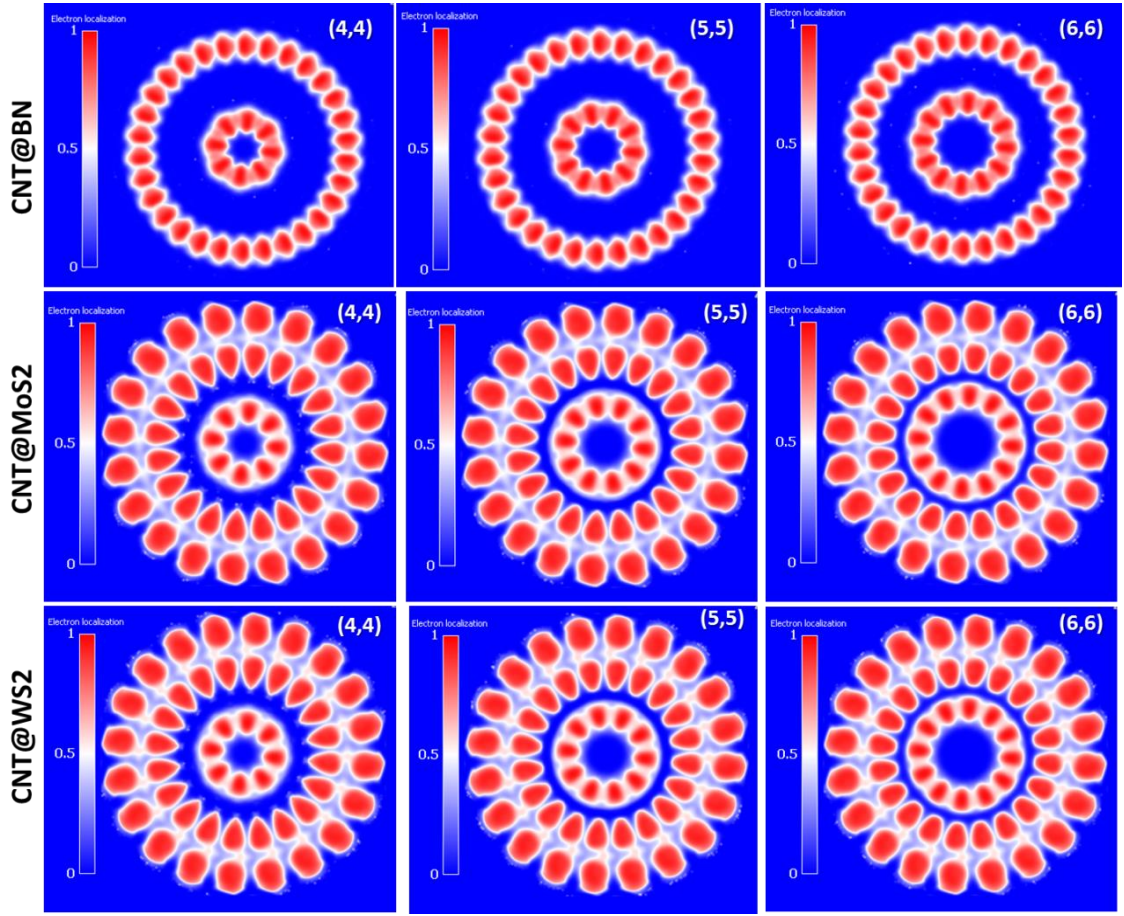

**Fig. 4:** The electron localization function plots in the x-y cut plane for the CNT@MS2 and CNT@BN core-shell nanotubes. (the type of core-shell NT is indicated on the left side and chirality on the individual tiles.)

A probabilistic measurement of where the electrons are concentrated in the structures, can be obtained from the plots of the electron localization function (ELF) shown in Fig. 4. In these plots the colorbar value 0 (blue) means zones without electrons, 1 (red) means areas with highest possibility of electron being located and 0.5 (white) represents an electron-gas like entity being present in the region. The ELF plots show a stronger more distinct localization of electrons near the C atom cores in case of CNT@MS2, than observed in the CNT@BN structures. In the $MS_2$ regions strong localization is seen around the chalcogen atoms with the inner S atoms having a distinct ovoid shaped appearance with the shorter half forming a tail-like perturbation towards the C atoms in the core of the tubes. As the CNT diameter increases with change in chirality, this "tail" tends to be suppressed. This is a result of the electron-electron repulsive forces taking more effect as the inner and outer tubes get closer, resulting in a stronger localization of the electrons on the atom cores, rather than the inter-tube region. W atoms have a more prominent electron-gas presence around them as compared to the Mo atoms. The presence of this electron gas is more concentrated near the metal atom cores than the nearby free space.

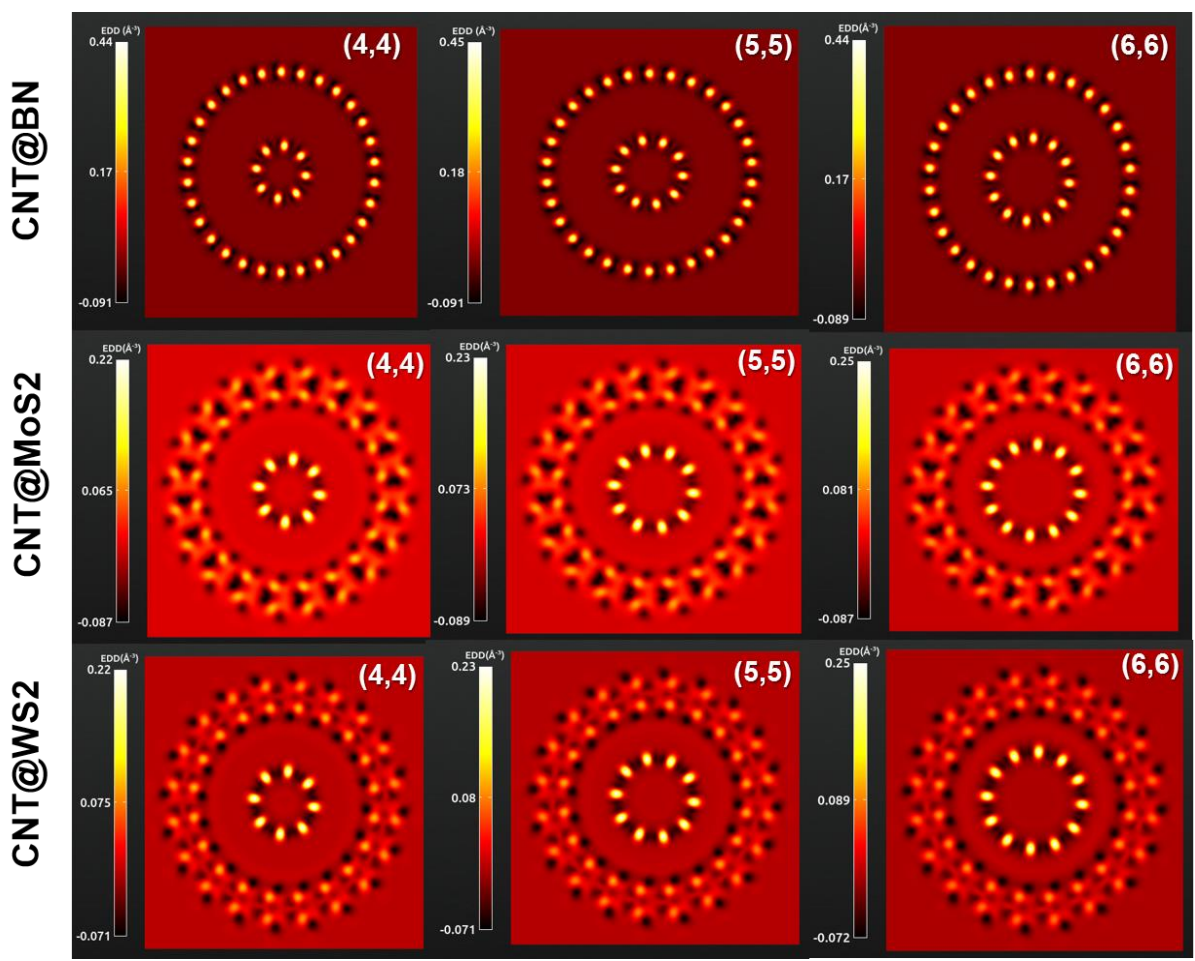

**Fig. 5:** The electron density difference (EDD) potential plots in the x-y cut plane for the CNT@MS$_2$ and CNT@BN core-shell nanotubes. (the type of core-shell NT is indicated on the left side and chirality on the individual tiles.)

The electron density difference, shown in Fig. 5, presents quite different pictures for CNT@BN and CNT@MS$_2$ systems. The former having an environment which is more towards the slightly negative values than the latter one. Strongly positive EDD on the atom cores of BN and C in the CNT@BN configuration suggest a tendency of increment of charge on atom cores and a stronger depletion of electron gas like presence in the inter-tube space in case of CNT@BN. For CNT@MS$_2$ the environment seems more towards a mildly positive EDD. The ability of chalcogen or carbon atoms to accumulate/pull charges from surroundings seem less strong as compared to that of the atom cores in CNT@BN. The EDD values at the carbon atoms in the CNT@BN is around 0.40-0.45 $Å^{-3}$, which is twice as strong as those in the carbon atoms in CNT@MS$_2$ (~ 0.20-0.25 $Å^{-3}$). The Mo atoms seem to have a more negative EDD as compared to W, therefore possibly resulting in a stronger dipole to be developed between the M-S pairs in MoS$_2$. More thorough understanding of the impact of this EDD distribution can be understood with the electrostatic difference potential.

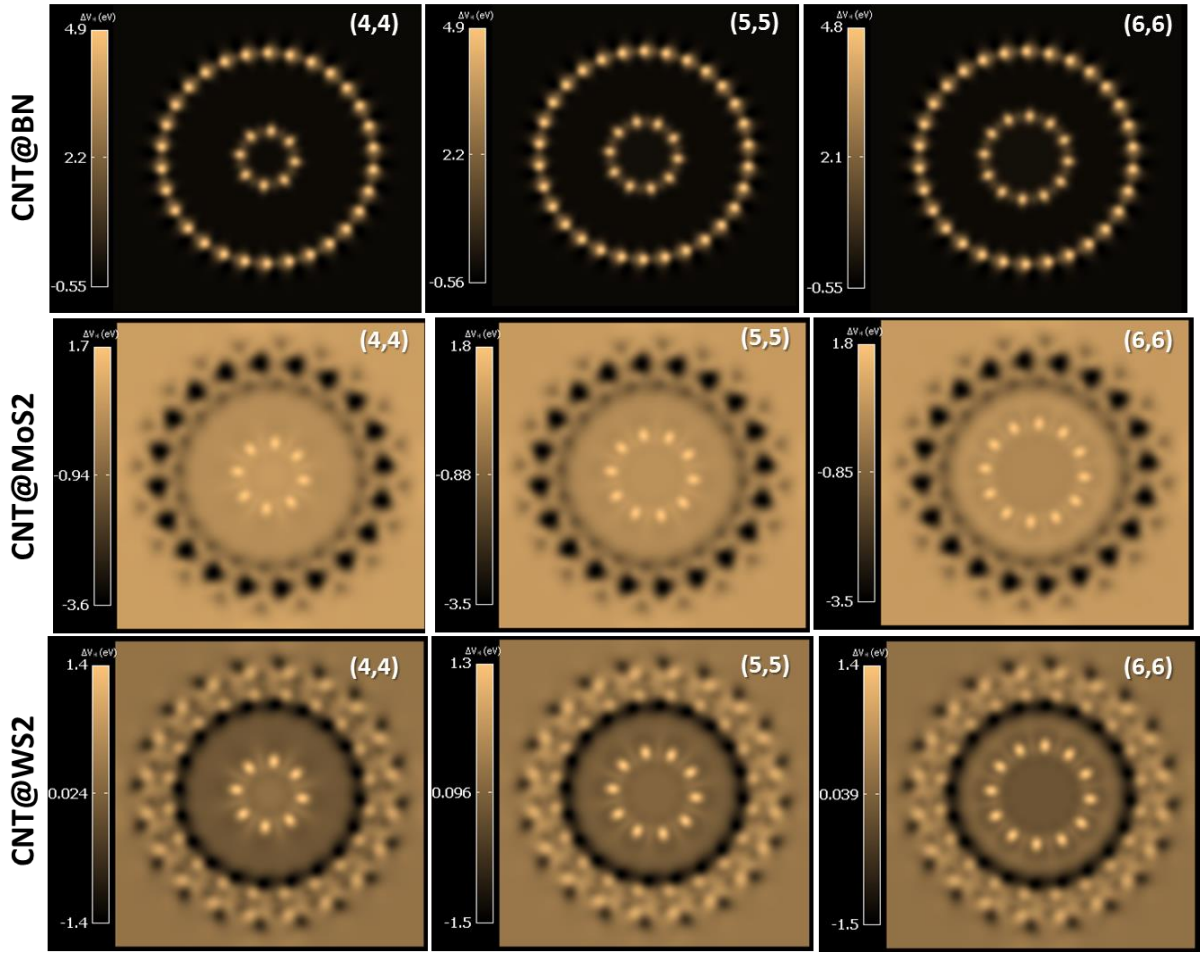

**Fig. 6:** The electrostatic difference potential plots in the x-y cut plane for the CNT@MS$_2$ and CNT@BN core-shell nanotubes. (the type of core-shell NT is indicated on the left side and chirality on the individual tiles.)

The electrostatic difference potential (EDP or $\Delta V_H$ ), shown in Fig. 6 is basically the potential developed due to the difference in the charge density due to superposition of the wavefunctions and the valence charge density that is calculated self-consistently. The EDP plots for the core-shell structures present a very contrasting view of the CNT@BN and the CNT@MS$_2$ configurations. For CNT@BN a highly positive potential of value around 4.8 eV is seen at the atom cores of C, B and N in a mostly negative potential environment. For CNT@MS$_2$ however the environment is mostly showing a mildly positive nature with the C atoms behaving as centres of relatively stronger positive (1.4 to 1.8 eV) potential. In case of CNT@WS$_2$ the potential due to the chalcogen atoms are more distinctly positive than for the CNT@MoS$_2$, although in both cases the potential on the chalcogen atoms is weaker than that on C atoms. For Mo atoms in the MoS$_2$ shell the potential drops significantly towards the negative (seen by alternating black patches), which is not so strong in case of the W atoms in CNT@WS$_2$. Overall the potentials in the CNT@MS$_2$ are quite lesser in magnitude compared to CNT@BN structures and tend more towards the negative region. The results in Fig. 5 and Fig. 6, thus represent a probability of charge transfer from the metal to the S atoms in the MS$_2$ shell. A region of potential inversion represented by a darker ring is also seen just before the inner S atoms (as we move radially outwards). The inversion which is the boundary of the potential screening effect of the CNT is more prominent for CNT@WS$_2$ than CNT@MoS$_2$.

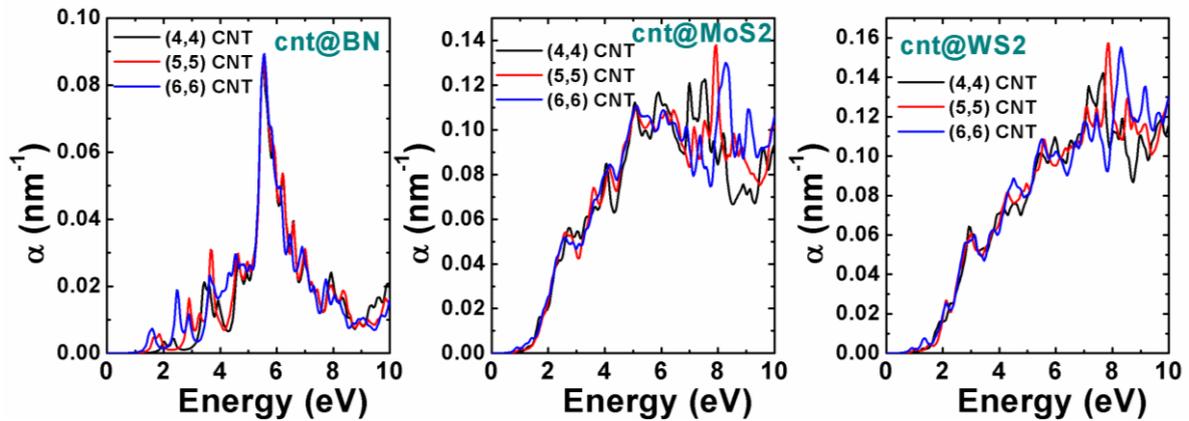

Fig.7: The simulated optical absorption spectra of for the CNT@MS$_2$ and CNT@BN core-shell nanotubes.

The optical absorption spectra (Fig. 7) shows absorption for the CNT@BN structures to be strong around 5.75 eV with minor absorption peaks at lesser energies of around 4 eV and between 2-3 eV depending on the chirality of the core CNTs. The variation in optical absorption with change in chirality is limited more or less to the minor peaks in the 1.5-4.5 energy range only. For the CNT@MS$_2$ structures the spectra is rather different in terms of the absence of a standout absorption peak of strength, rather a number of peaks of similar strength is observed around the 6-9 eV energy range for the CNT@MS$_2$ structures. However the maximum magnitude of absorption coefficient in CNT@BN is lesser than those corresponding to the CNT@MS$_2$ structures. Chirality of the CNT plays a more significant part here than in the CNT@BN core-shell tubes, with a clear blue shift of the absorption maxima being observed with increase in the chirality of the CNT for both the CNT@MoS$_2$ and CNT@WS$_2$ tubes.

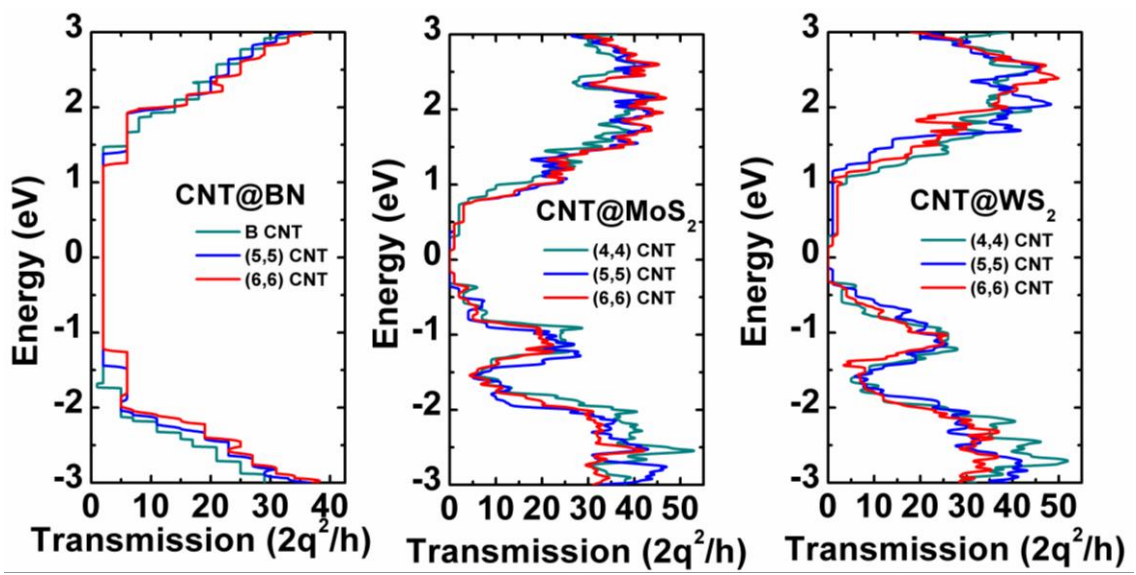

Fig. 8: The simulated conductance of for the CNT@MS$_2$ and CNT@BN core-shell nanotubes.

The transmission characteristics of the core-shell structures (Fig. 8), shows that the CNT@MS$_2$ structures to be more conductive than the CNT@BN structures. CNT@BN also shows a metallic behaviour in terms of a uniform non-zero transmission observed from -1.75 to 1.5 eV (including the fermi level $\varepsilon_F$). This is different from CNT@MS$_2$, which possess a small number of localized states near the fermi energy (as reflected in the DOS plots) but they do not contribute to the transmission as such. The difference in chirality does not seem to affect the transmission in CNT@BN very much, with only minor changes. For CNT@MS$_2$ it is seen that with increasing For CNT@MS$_2$ the transmission is not metallic in nature but the number of transmission states present nearer to the Fermi level is much larger. This is likely to be more useful for switching applications where a minor change in the applied bias could bring in a larger number of states within the bias window (and thereby creating a more significant current difference). The staircase like behaviour in case of the CNT@BN is not present as much in CNT@MS$_2$ tubes. A reason behind this can be the larger difference in the band gap between the constituent materials, resulting in less overlap of the bands (especially at lower energies, as also seen in the DOS plots) between the core and shell tubes. This nature ensures that the combined system (i.e. core-shell) retains more of the pristine transmission properties of the individual components and therefore a larger degree of staircase.

**IV. CONCLUSION**

In this paper we carry out atomistic simulation of core-shell structures of *(n,n)* metallic CNTs with larger diameter armchair MS$_2$ and BN nanotubes to study their electronic, optical and carrier transport properties. The LDA-DFT simulations for the material properties and the DFT-GF studies carried out for the electron transmission spectra of CNT@MS$_2$/BN gives interesting details about the fundamental electrostatics of the systems and its optical and quantum transport behavior. The MS$_2$ shells seem to open up a moderate indirect band gap in the core-shell tubes while the BN encapsulation doesn't affect the original metallic nature of the CNT. Greater tuning of the carrier effective mass for varying chirality of CNT is observed in case of the CNT@MS$_2$ structure. The Mo atoms seem to have higher electron density around them compared to W atoms, while the overall density is much more localized near the atom cores and stronger in magnitude for the CNT@BN. Also the W atoms show a more prominent electron-gas presence around them than Mo atoms as found in the electron localization functions. In CNT@BN a highly positive potential in the atom cores is seen in a overall negative environment of electrostatic difference potential (EDP), while the environment seems mildly positive EDP with atom cores having more positive potential in the MS$_2$ encapsulated systems. In terms of optical absorption a strong and sharper peak is observed around 6 eV for the CNT@BN compared to a more broad absorption spectra of the CNT@MS$_2$. The transmission for CNT@BN clearly shows a metallic nature suited for interconnect applications, while CNT@MS$_2$ shows non-metallic transport although with a larger number of transmission states near fermi level, which could be useful for switching applications. Overall the tuning of the electrostatics, optical and transport properties of CNTs by encapsulation in different shells ($MS_2$ or BN nanotubes) seems to open up good prospects for the use of these structures in nanoelectronics and photonics.

**ACKNOWLEDGEMENT**

The work is supported by the Hanse-Wissenschaftskolleg (HWK) fellowship in energy research 2016-17. HPC cluster facility with ATK on-demand, provided by Sabalcore Computing Inc. was utilized for calculations. Part of the work was carried out at Indian Institute of Engineering Science and Technology, Shibpur with DST INSPIRE Faculty Grant No. IFA-13 ENG-62.